# NMR in magnetic molecular rings and clusters

F. Borsa[a,b], A. Lascialfari[a] and Y. Furukawa[c]

a) Dipartimento di Fisica "A.Volta" e Unita' INFM , Universita' di Pavia, 27100 Pavia, Italy
b) Department of Physics and Astronomy and Ames Laboratory, Iowa State University, Ames, IA 50011
c) Division of Physics, Graduate School of Science, Hokkaido University, Sapporo 060-0810, Japan

**Contents**

# I. Introduction

In the last years there has been a great interest in magnetic systems formed by a cluster of transition metal ions covalently bonded via superexchange bridges, embedded in a large organic molecule [1-5]. Following the synthesis and the structural and magnetic characterization of these magnetic molecules by chemists, the physicists realized the great interest of these systems as a practical realization of zero-dimensional model magnetic systems. In fact the magnetic molecules can be synthesized in crystalline form whereby each molecule is magnetically independent since the intramolecular exchange interaction among the transition metal ions is dominant over the weak intermolecular, usually dipolar, magnetic interaction.

Magnetic molecules (see Figs. 0 (a)-(d) for some example systems) can be prepared now a day with an unmatched variety of parameters: (i) the size of the magnetic spin can be varied spanning from high "classical" spins to low "quantum" spins by using different transition metal ions i.e. $Fe^{3+}$, $Mn^{2+}$ (s=5/2) ; $Mn^{3+}$ (s=2) ; $Cr^{3+}$,$Mn^{4+}$ (s=3/2); $Cu^{2+}$, $V^{4+}$(s=1/2) ; (ii) the exchange interaction can go from antiferromagnetic (AFM) to ferromagnetic (FM) with values of the exchange constant J ranging from a few Kelvins to more than 1000K ; (iii) the geometrical arrangement of the magnetic core of the molecule can be as simple as a coplanar regular ring of magnetic ions as found in many Fe and Cr rings to totally asymmetric three dimensional clusters such as $[Fe_8(N_3C_6H_{15})_6O_2(OH)_{12}]^{8+}\cdot[Br_8\cdot 9H_2O]^{8-}$ (in short Fe8) ; (iv) the symmetry of the magnetic Hamiltonian can go from isotropic Heisenberg type as in most cases to easy axis or easy plane. Choosing from this variety of model systems one can investigate fundamental problems in magnetism taking advantage of the fascinating simplicity of zero-dimensional systems. Examples of issues of interest are the transition from classical to quantum behavior, the effect of geometrical frustration, the form of the spectral density of the magnetic fluctuations, the spectrum of the low- lying excitations with the connected problem of quantum spin dynamics and tunneling.

NMR has proved to be a powerful tool to investigate both static and dynamic properties of magnetic systems. In particular it has been very successful in addressing some special features in low dimensional magnetic systems. For example in one dimensional magnetic Heisenberg chains the long-time persistence of spin correlation has dramatic consequences on the field dependence of the nuclear spin-lattice relaxation rate $T_1^{-1}$ which directly probes the low-frequency spectral weight of spin fluctuations. From $T_1$ and $T_2$ measurements one can detect the crossover of spin dimensionality from Heisenberg to XY to Ising as a function of temperature in both one and two dimensional systems. One final example among the many is the study of the gap in energy in Haldane s=1 spin chains.

With the above scenario in mind we have undertaken a systematic NMR investigation of molecular nanomagnets since back in 1996. The present review tries to give an account of the main results obtained so far and of the many exciting projects that still lie ahead. The work was done through a continuous very fruitful collaboration among three NMR laboratories: at the University of Pavia, Italy, at Iowa State University and Ames Laboratory, Ames, IA, USA and at Hokkaido University, Sapporo, Japan with occasional very useful collaborations with the high field NMR lab. in Grenoble, France. None of the work could be done without the precious collaboration and help of our colleagues in chemistry at the University of Florence and of Modena, Italy and at Ames Laboratory in USA who synthesized and characterized the samples used in the NMR work.

In Section II we will mention some of the problems encountered in doing NMR in molecular nanomagnets. These are related for example to the presence of many non equivalent nulclei, to very broad and structured resonance lines, to very short relaxation rates, to vanishingly weak signals. In the three following sections (III, IV, V) we present and discuss the experimental data. To organize the considerable amount of data we chose somewhat arbitrarily to divide them according to the temperature range. In fact the physical issues regarding the magnetic properties and the spin dynamics of the molecular nanomagnets depend on the relative ratio of the thermal energy $k_B T$ and the magnetic exchange energy J. At high T the individual magnetic ions in the molecule behave as weakly correlated paramagnetic ions; at very low T the individual spins are locked into a collective quantum state of total spin S; at intermediate T the interacting spins develop strong correlations in a way similar to what happens in magnetic phase transitions in three dimensional systems. In the illustration of the physical issues encountered in the different temperature ranges we utilize the most representative results for different kind of molecules. The magnetic molecules which were investigated by NMR but whose results are not mentioned in Sections III, IV and V are reviewed separately in Section VI.

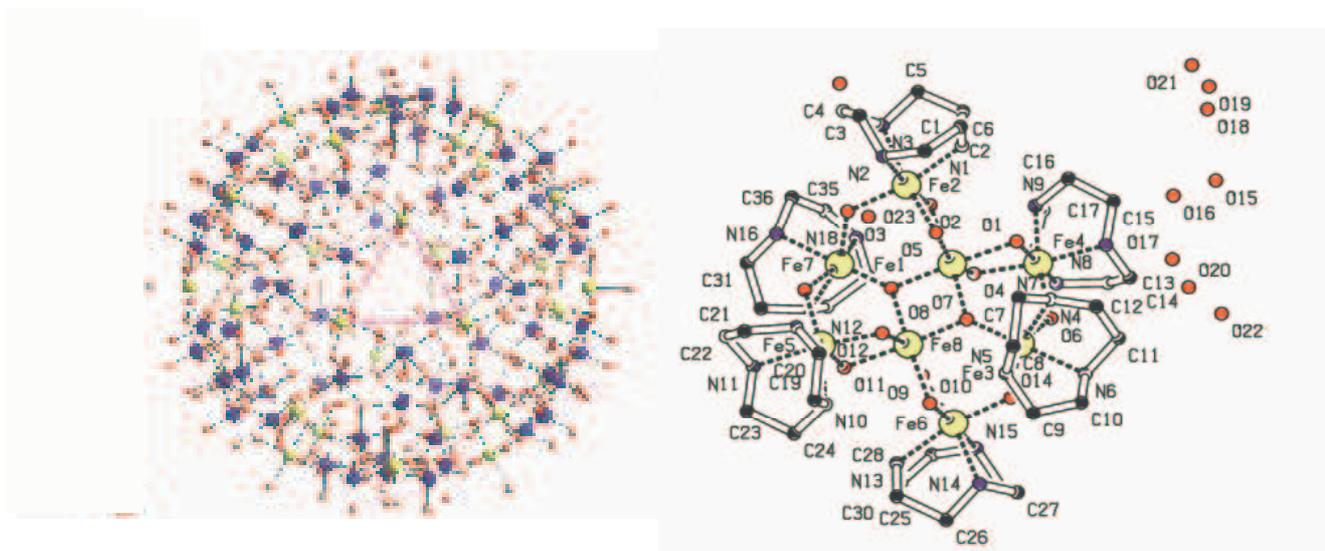

Fig. 0(a)  Structure of Fe30 [2]. Yellow : iron , blue : molybdenum , red : oxygen. Hydrogen atoms are not shown for simplicity.

Fig. 0(b)  Structure of Fe8 [1]. The hydrogen atoms are not shown for simplicity.

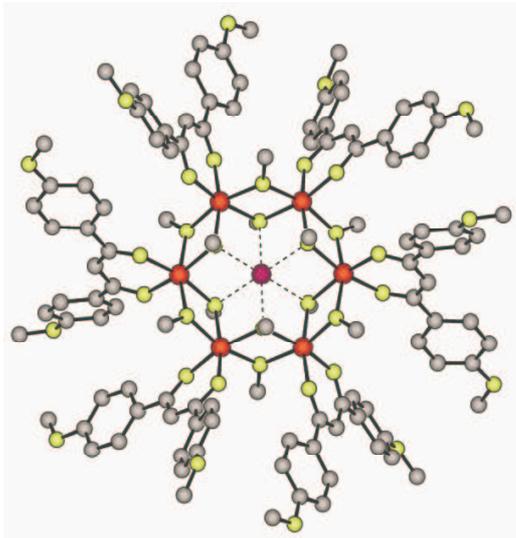 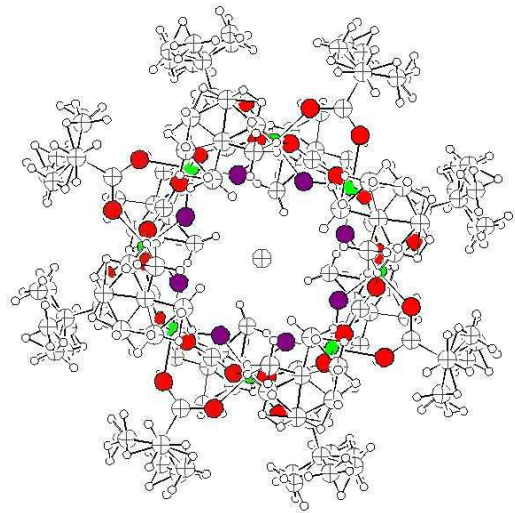

Fig. 0 (c) Structure of Fe6(X) (X= alkaline ion). Red : iron , yellow : oxygen , grey : carbon , violet : alkaline atom (Li,Na). Hydrogen atoms omitted for clarity.

Fig. 0(d) Structure of Cr8. Main atoms : green : chromium, black : fluorine , red : oxygen.

## II. Challenges of NMR in molecular nanomagnets

Molecular nanomagnets offer a wide variety of nuclei which can be used to probe the magnetic properties and the spin dynamics. Most of the measurements were done on proton NMR. In this case the signal is very strong but the width of the spectrum and the presence of many inequivalent protons in the molecule require some special attention in the analysis of the results. Due to the above reasons the recovery of the nuclear magnetization was found in many cases to be strongly non exponential. There are two sources for non exponential recovery. The first is due to incomplete saturation of the broad NMR spectrum which leads to an initial fast recovery of the nuclear magnetization due to spectral diffusion. If the spectrum is not too wide (at most twice the spectral width of the rf pulse) one can still saturate the whole line by using a sequence of rf pulses provided that the condition of $T_1$ much longer and $T_2$ much shorter than the pulse spacing can be met. In nanomagnets with magnetic ground state like $[Mn_{12}O_{12}(CH_3COO)_{16}(H_2O)_4]$ (in short Mn12) and Fe8 [1] the proton spectrum can be as wide as 4MHz and structured at low temperature. In this case the spectrum has to be acquired by sweeping the field and/ or the frequency and plotting the amplitude of the echo signal after proper correction for changes of irradiation conditions. In this case the relaxation can only be measured at given points of the spectrum and the spectral diffusion effect cannot be avoided. One can try to establish the percentage of the fast initial recovery which is affected by spectral diffusion and exclude that from the measurement but a large systematic error can still be unavoidable. The second source of non exponential recovery is due to the presence in the molecule of protons having a different environment of magnetic ions and thus having a different relaxation rate. If $T_2$ is fast compared to $T_1$ then a common spin temperature is achieved during the relaxation process and the recovery is exponential with a single $T_1^{-1}$ which is a weighted average of the rates of the inequivalent protons in the molecule. In the opposite limit encountered when $T_1$ is comparable to $T_2$ each nucleus or group of nuclei relax independently with its own spin temperature and the recovery of the nuclear magnetization results in the sum of exponentials:

$$n(t) = [ M(\infty) - M(t) ]/ M(\infty) = \Sigma_i\, p_i\, \exp(-t/T_{1i}) \qquad (1)$$

If there is a continuous distribution of $T_1$'s the recovery follows a stretched exponential function $\exp(-(t/T_1^*)^\beta)$ where $\beta < 1$ is the smaller the wider is the distribution and $T_1^*$ is a relaxation parameter related to the distribution of $T_1$'s in a non trivial manner. When the recovery is non exponential it is best to measure the $T_1$ parameter from the recovery of the nuclear magnetization at short times. In fact the slope at $t \to 0$ of the semilog plot of n(t) vs t yields an average relaxation rate $T_1^{-1} = \Sigma_i\, p_i\, \exp(-t/T_{1i})$. Unfortunately, in most cases the situation is intermediate between the two above limiting cases. In this circumstance there is no simple way that one can define a spin-lattice relaxation parameter. Since in many instances one is interested in the relative changes vs T and H one can simply define an effective relaxation parameter R by taking the time at which the recovery curve n(t) reduces to 1/e of the initial value.

Other isotopes which have been utilized for NMR studies of magnetic molecules include $^2$H, $^{13}$C, $^7$Li, $^{23}$Na, $^{63,65}$Cu. The disadvantage of a weaker signal in $^{13}$C is in part compensated by the advantage of having a nucleus with strong hyperfine coupling to the magnetic ions and with less number of inequivalent sites with respect to protons. For the remaining quadrupole nuclei there is the additional information obtained by the quadrupole coupling with the electric field gradient. When the quadrupole interaction is sufficiently strong to remove the satellite transition from the central line the non exponential decay of the nuclear magnetization becomes very difficult to analyze because besides the non equivalent sites one has to take into account the intrinsic non exponential decay due to unequal separation of the Zeeman levels. The $^{63,65}$Cu case is the only one where, to our knowledge, a pure NQR experiment has been performed in molecular clusters ( i.e. [Cu$_8$(dmpz)$_8$(OH)$_8$] · 2C$_6$H$_5$NO$_2$ , in short Cu8). The NQR spectrum was found to contain several lines in the frequency range 16-21MHz.

Very useful information were obtained from the $^{55}$Mn and $^{57}$Fe NMR in Mn12 and Fe8 clusters respectively. The NMR of the above nuclei can be observed only at low temperature (T< 4K) since with increasing temperature the relaxation times $T_1$ and $T_2$ become too short. The $^{55}$Mn and $^{57}$Fe (in isotopicaly enriched sample) NMR were detected both in zero field and in an externally applied field . The zero field $^{55}$Mn NMR spectrum in Mn12 consists of three quadrupole broadened lines (i.e., each several MHz wide) in the frequency range 230-370MHz while the $^{57}$Fe NMR spectrum in Fe8 is made of eight different lines rather narrow (i.e., 100kHz) in the frequency range 63-73 MHz. Both Mn12 and Fe8 are ferrimagnetic molecules at low temperature. However, since there are no domain walls and the anisotropy is very high no signal enhancement due to the rf enhancement in domain walls and/or domains is present contrary to normal ferro or ferri magnetic long range ordered systems. As a consequence the NMR signal intensity in zero external field is small (particularly in Fe8) even at low temperature since the frequency range of the overall spectrum is quite broad.

## III. NMR at high temperature ( $k_BT \gg J$)

Most of the magnetic molecular clusters investigated are characterized by exchange constants J which are well below the room temperature energy value $k_BT$. Exceptions to this is the Cu8 ring and to a certain extent also Mn12 and Fe8 clusters. If $k_BT \gg J$, the magnetic moments in the cluster are weakly correlated and the system behaves like a paramagnet at high temperature. In this case the nuclear spin lattice relaxation due to the coupling to the paramagnetic ions should be field independent as indeed found in paramagnets such as $MnF_2$. On the other hand the $T_1$ in molecular nanomagnets is strongly field dependent as shown for a number of systems in Figs. 1(a)-(f). All data reported here refer to proton NMR. The field dependence of $T_1$ is a characteristic feature of the zero dimensionality of the magnetic system. A similar field dependence is well known to occur in one dimensional magnetic chains and, to a lesser extent, in two dimensional paramagnets. The fundamental reason for this is that in Heisenberg isotropic paramagnets the time dependence of the spin correlation function has a long time persistence in low dimensions. We will review briefly this result in the following.

In the weak-collision approach $T_1^{-1}$ can be expressed as [6]:

$$T_1^{-1} \propto \Sigma_{ij} \alpha_{ij} J_{\pm}^{ij}(\omega_e) + \Sigma_{ij} \beta_{ij} J_z^{ij}(\omega_L) \quad (2)$$

where i,j number the electronic spins, $\omega_e$ and $\omega_L$ are the Larmor frequencies of the electron and of the nucleus respectively, $\alpha_{ij}$ and $\beta_{ij}$ are geometrical factors and $J_{\pm,z}^{ij}$ are the transverse and longitudinal spectral densities of the spin fluctuations. In Eq. 2, $J_{\pm,z}^{ij}(\omega)$ can be expressed by the Fourier transform (FT) of the spin correlation function (CF):

$$J_{\pm,z}^{ij}(\omega) = \int G_{ij}^{\alpha}(r,t) \exp(i\omega t) \, dt \quad (3)$$

An approximate expression for the correlation function can obtained for an infinite Heisenberg classical chain at high temperature by matching the short time expansion to the long time diffusive behavior due to the conservation of the total spin and of its component in the direction of the applied field [7]. For temperatures $T \gg J/k_B$ the conservation property can be incorporated for spins on a ring by means of a discretized diffusion equation to which cyclic boundary conditions are applied. For this model it is found [8,10] that the auto-correlation function (CF) decays rapidly at short times until it reaches a constant value which depends on the number of spins in the cluster. The plateau in the CF is reached after a time of the order of $10\omega_D^{-1}$ where $\omega_D$ is the exchange frequency given at the simplest level of approximation by [6]:

$$\omega_D = (2\pi J/h)[S(S+1)]^{1/2} \quad (4)$$

with J the exchange constant between nearest neighbor spins S. The same result is obtained for the CF by using a one dimensional hopping model on a closed loop [9] or by calculating the spin correlation function with a mode-coupling approach [16] The leveling off of the time dependence of the CF at a value approximately given by 1/N with N the number of spins in the cluster is the result of the conservation of the total spin component for an isotropic spin-spin interaction. In practice the anisotropic terms in the spin Hamiltonian will produce a decay of the CF via energy exchange with the "lattice".

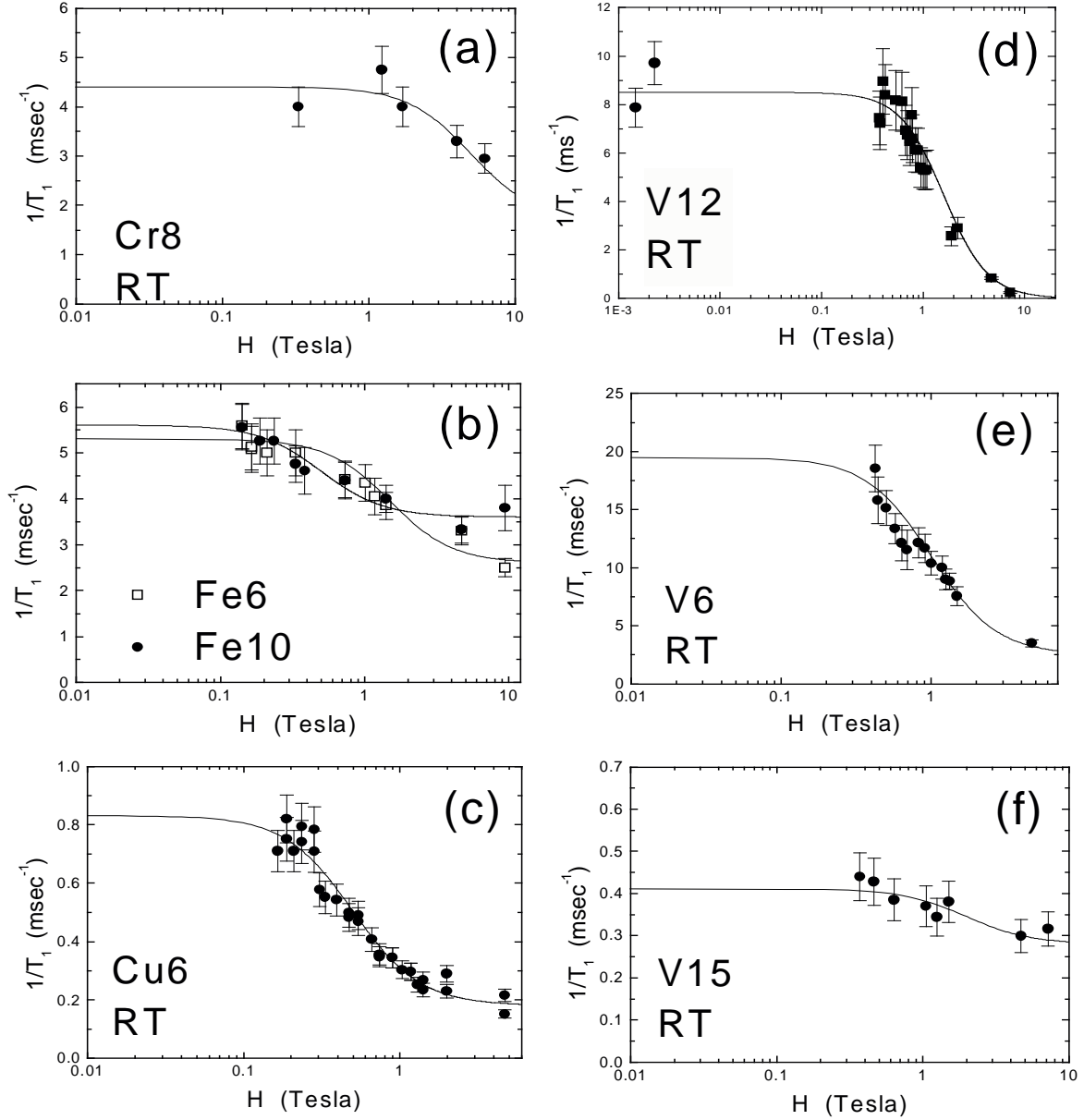

Fig. 1 : The external field dependence of $^1$H-$1/T_1$ at room temperature. (a) Cr8; (b) Fe6 and Fe10; (c) Cu6; (d) V12; (e) V6; (f) V15. The solid lines in the figure are calculated results using equation (7) with a set of parameters listed in Table I. All samples are in powder form. The two points at very low field in Fig.(d) for V12 refer to spin-lattice relaxation rates in the rotating frame, $T_{1\rho}$, measured at 4.7 Tesla.

A sketch of the time decay of the CF and of the corresponding spectral density is shown in Fig. 2. The decay at long time of the CF has a cut-off at a time $\Gamma_A^{-1}$ due to the anisotropic terms in the spin hamiltonian [11,16]. In the following we will discuss the magnetic field dependence of the nuclear relaxation rate at room temperature in terms of a simplified model which incorporates the theoretical understanding of the spin dynamics in clusters as described above.

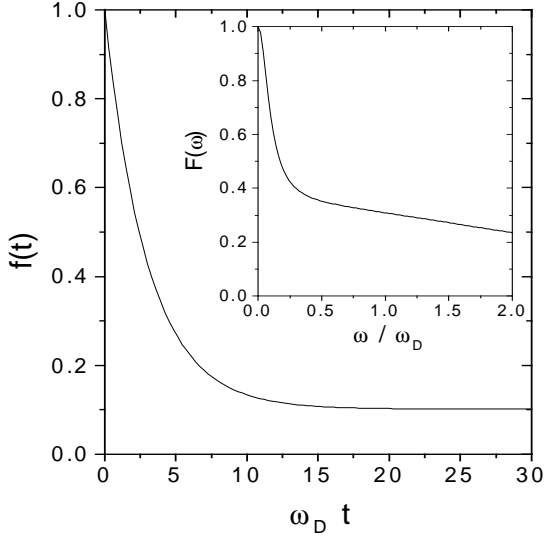

Fig. 2 : Sketch of the decay in time of the autocorrelation function of the spins on magnetic ions. The initial fast decay is characterized by the constant $\Gamma_D^{-1}$ while the slow decay at long time is characterized by the constant $\Gamma_A^{-1}$. In the inset the behavior of the Fourier transform i.e. the spectral density is shown.

On the basis of the time dependence of the CF discussed above and sketched in Fig. 2, we model the spectral function in Eq. 2 as the sum of two components [12]:

$$\Phi^\pm(\omega)=\Phi^z(\omega)=\Phi'(\omega)+\Gamma_A/(\omega^2+\Gamma_A^2)=\Gamma_D/(\omega^2+\Gamma_D^2)+\Gamma_A/(\omega^2+\Gamma_A^2) \quad (5)$$

where we assume the same CF for the decay of the transverse ($\pm$) and longitudinal (z) components of the spins. The first term in Eq.5 represents the Fourier transform (FT) of the initial fast decay of the CF while the second term represents the FT of the decay at long time of the CF due to anisotropic terms in the spin Hamiltonian and we model this second part with a Lorenzian function of width $\Gamma_A$.

From Eq.4 one can estimate that the exchange frequency $\omega_D$ is of the order of $10^{13}$ Hz for typical values of $J/k_B$ (10-20K) and spin values S (1/2-5/2). The spectral function $\Phi'(\omega)$ in Eq.5 reaches a plateau and becomes almost frequency independent for $\omega < \omega_D/10$. For the magnetic field strength used in the experiment (see Fig. 2) both $\omega_n$ and $\omega_e$ are smaller than $\omega_D/10$. Thus we will assume $\Phi'(\omega_n) = \Phi'(\omega_e) = \Gamma_D^{-1}$ in Eq. 5 where the characteristic frequency $\Gamma_D$ is of the same order of magnitude as $\omega_D/10$. Finally, by assuming $\omega_n << \Gamma_A$ in Eq. 5, Eq. 2 can be rewritten as:

$$1/T_1 = K\ [1/2 A^\pm (\ \Gamma_A/(\omega_e^2 + \Gamma_A^2)+1/2\ A^\pm/\Gamma_D + A^z (1/\Gamma_D + 1/\Gamma_A)] \quad (6)$$

where the constants $A^\pm$ and $A^z$ are averages over all protons in the molecule of the products of the hyperfine dipolar tensor components $\alpha_{ij}$ and $\beta_{ij}$ respectively (see Eq. 2). The constant K which has been factored out from the dipolar tensor coefficients is given by $K=(h\gamma_n\gamma_e)^2/(4\pi) = 1.94 \cdot 10^{-32}$ ($sec^{-2}cm^6$ ). The width $\Gamma_A$ of the narrow component in the spectral function represents the frequency which characterizes the exponential time decay of the spin CF in the cluster due to anisotropic terms in the spin Hamiltonian.

The experimental data in Figs. 1(a)-(f) were fitted by using an expression of the form of Eq. 6.

$$1/T_1 = A/(1 + (H/B)^2) + C \quad (msec^{-1}) \quad (7)$$

where the magnetic field H is expressed in Tesla and $B = \Gamma_A / \gamma_e$ (Tesla). The fitting parameters for the different rings and clusters are summarized in Table I.

TABLE I. The fitting parameters in Eq. 7 for the different molecular rings and clusters (see text for complete chemical formula), in powders.

| Single molecular magnet | A ($msec^{-1}$) | B (Tesla) | C($msec^{-1}$) |
|---|---|---|---|
| Cr8 (AFM ring- s=3/2)) | 2.7 | 5 | 1.7 |
| Fe6 (AFM ring-s=5/2)) | 2.7 | 1.5 | 2.6 |
| Fe10 ( AFM ring-s=5/2)) | 2 | 0.5 | 3.6 |
| Cu6 (FM ring-s=1/2) | 0.65 | 0.5 | 0.18 |
| V12 (AFM square-s=1/2) | 8.5 | 1.6 | ~0 |
| V6 (AFM triangle-s=1/2) | 17 | 1 | 2.5 |
| V15 ( AFM ring-s=1/2) | 0.13 | 2 | 0.28 |

The most significant parameter in Table I is B which measures the cut off frequency $\Gamma_A$ of the electronic spin-spin correlation function. Except for the Cr8 case (complete formula : [$Cr_8F_8Piv_{16}$], Hpiv=pivalic acid) B is around 1 Tesla corresponding to $\Gamma_A \approx 10^{11}$ rad $sec^{-1}$ or $h\Gamma_A/k_B \approx 1$ K .The cut-off effect is provided, in principle, by any magnetic interaction which does not conserve the total spin components. In practice, such small terms stem from a variety of mechanisms including intracluster dipolar and anisotropic exchange interaction, single ion anisotropies, interring dipolar or exchange interactions etc..[11]. A detailed calculation for Fe6 based on intraring dipolar interaction yielded $\Gamma_A = 1.5 \ 10^{11}$ $sec^{-1}$ [16] . A similar estimate for Cu6 based on known anisotropic nearest neighbor ( exchange and dipolar) contributions to nearest neighbor interactions yielded $1.4 \ 10^{11}$ $sec^{-1}$ [16]. Both these results can account very well for the experimental findings in Table I . From the comparison of Eqs. 6 and 7 one has $A= K A^{\pm}/2\Gamma_A$ and $C \approx K A^z/\Gamma_A$ (since $\Gamma_D \gg \Gamma_A$). Thus the order of magnitude of the hyperfine constants is $A^{\pm} \approx A^z \approx 1 \div 10 \ 10^{46}$ $cm^{-6}$. Since $A^{\pm}$, $A^z$ are the product of two dipolar interaction tensor components they are of order of $r^{-6}$ where r is the distance between a $^1H$ nucleus and a transition metal local moment. For most of the rings the value of the hyperfine constants is consistent with a purely nuclear-electron dipolar interaction.

For V12 (complete formula : $(NHEt)_3[V^{IV}_8V^V_4As_8O_{40}(H_2O)]\cdot H_2O$) and V6 (complete formula of one variant : $Na_6[H_4(V_3L)_2P_4O_4]\cdot 18H_2O$ ), the $A^\pm$ ($\gg A^z$) hyperfine constant is one order of magnitude higher indicating the presence of an additional contribution probably due to a contact interaction due to the admixing of the hydrogen s wave function with the d wave function of the Vanadium ions. An alternative way to explain the anomalous values (A>>C ) for V6 and particularly for V12 in Table I is to go back to Eq. 7 and assume that $B = \Gamma_A/\gamma_N$ instead of $B = \Gamma_A/\gamma_e$ . This implies that the cut-off frequency $\Gamma_A$ is much less than in other clusters namely of order of the nuclear Larmor frequency in Eq.6. In this case the value of the constant C in Eq.7 is close to zero in agreement with the experiments as can be seen easily by modifying in the appropriate way the approximate Eq.6. It is, however, difficult to justify such small value for the cut-off frequency in V12 [16b].

## IV. NMR at intermediate temperatures ( $k_B T \approx J$)

As the temperature is lowered and it becomes comparable to the magnetic exchange interaction J strong correlations in the fluctuations of the magnetic moments of the molecule start building up. The situation is analogous to macroscopic three dimensional magnetic systems when the temperature approaches the critical temperature for the transition to long range magnetic order. In molecular magnets, as a result of the finite size of the system the low lying magnetic states are well separated among themselves. Therefore the correlation of the magnetic moments at low temperature has to be viewed as the result of the progressive population by the magnetic molecule of the collective low lying quantum total spin states without any phase transition. In this intermediate temperature range two very interesting situations arise which can be investigated by NMR and relaxation. On one hand one can follow the evolution of the electronic spin correlation function as the system crosses over from an uncorrelated finite size paramagnet to a total spin S collective quantum state. On the other hand one can investigate the nature of the fluctuations of the local electronic spin while the system is in its ground state but at a temperature for which excitations to higher states are important. We will consider in the following examples of the two situations. For the evolution of the spin correlation function we will refer to simple antiferromagnetic rings having a total spin S= 0 ground state. For the thermal fluctuations in the ground state we will refer to ferromagnetic clusters having a total high spin S=10 ground state.

In order to describe the nuclear spin-lattice relaxation in a magnetic system in presence of a correlated spin dynamics it is more convenient to express the nuclear $T_1$ in terms of the q components of the electronic spins [6,17,18]:

$$1/T_1 = (h\gamma_n\gamma_e)^2/(4\pi) \int dt \cos(\omega_n t) \int dq \, (1/4 \, A^{\pm}(q) <S^{\pm}_q(t) S^{\pm}_{-q}(t)> \\ + A^z(q) ) <S^z_q(t) S^z_{-q}(t)>) \qquad (8)$$

or in terms of the response functions by using the fluctuation- dissipation theorem [18b]:

$$1/T_1 = (h\gamma_n\gamma_e)^2/(4\pi g^2\mu_B^2) \, k_B T \, [1/4\Sigma_q A^{\pm}(q) \chi^{\pm}(q) f_q^{\pm}(\omega_e) + \Sigma_q A^z(q)\chi^z(q)f_q^z(\omega_n)] \qquad (9)$$

where $\gamma_n$ and $\gamma_e$ are the gyromagnetic ratios of the nucleus and of the free electron respectively, g is the Lande's g factor, $\mu_B$ is the Bohr magneton, $k_B$ is the Boltzmann constant. The coefficients $A^{\pm}(q)$ and $A^z(q)$ are the Fourier transforms of the spherical components of the product of two dipole-interaction tensors [describing the hyperfine coupling of a given proton to the magnetic moments] whereby the symbols ± and z refer to the components of the electron spins transverse and longitudinal with respect to the quantization direction which is here the external magnetic field . The collective q-dependent spin correlation function is written as the product of the static response function times a normalized relaxation function $f_q^{\pm,z}(\omega)$.

At high temperature ($k_B T >> J$) one can neglect in Eq. 9 the q-dependence of the generalized susceptibility $\chi^\alpha(q)$ and of the spectral density function $f_q^\alpha(\omega)$. If one assumes an isotropic response function $1/2 \, \chi^{\pm}(q) = \chi^z(q) = \chi(q=0)$ and one takes a q-independent average value for the dipolar hyperfine interaction of the protons with the local moment of the electronic spins : $A^{\pm}(q) = A^{\perp}$ ; $A^{z}(q) = A^{z}$ in units of cm$^{-6}$, then Eq. 9 reduces in this high temperature limit to :

$$1/T_1 = (h\gamma_n\gamma_e)^2/(4\pi g^2 \mu_B^2) \; k_B T \; \chi(q=0) \; [\; 1/2 \; A^{\pm}\Phi^{\pm}(\omega_e) + A^z \Phi^z(\omega_n) \;] \qquad (10)$$

If one further assumes for the spectral density of the spin correlation the expressions Eq. 5 then Eq. 10 reduces to Eq. 6 used in the previous paragraph to analyze the field dependence of $1/T_1$ at room temperature. By decreasing the temperature at values such that $k_BT$ becomes comparable to J one expects that the nuclear spin lattice relaxation rate displays a characteristic temperature dependence related to the correlated spin dynamics according to Eqs. 8 and 9.

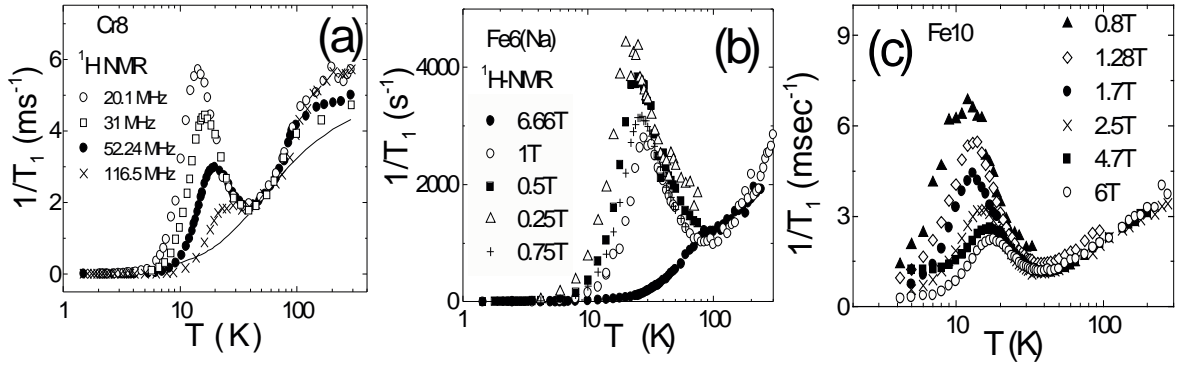

Fig. 3 : Temperature dependence of $^1$H-$1/T_1$ under various external magnetic fields. (a) Cr8, (b) Fe6(Na) and (c) Fe10. The solid line in (a) shows the temperature dependence of $\chi T$ in arbitrary unit.

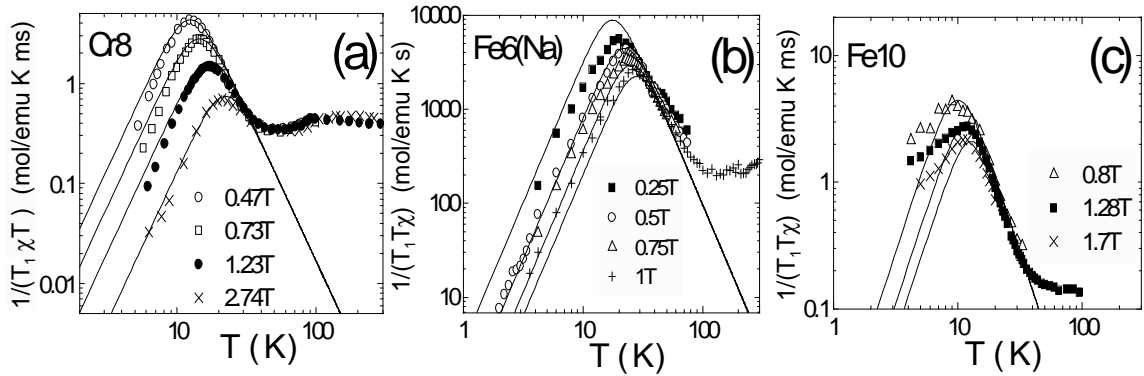

Fig. 4 : $1/(T_1\chi T)$ as a function of temperature in log-log plot. (a) Cr8; (b) Fe6(Na); (c) Fe10. The solid lines in the figures are theoretical calculation (see text). All samples are in powders.

## a) Thermal fluctuations in AFM rings

Measurements of proton $T_1$ as a function of temperature in a number of antiferromagnetic molecular rings has shown a surprisingly big enhancement of the relaxation rate at low temperatures resulting in a field dependent peak of $T_1^{-1}$ centered at a temperature of the order of the magnetic exchange constant $J/k_B$ ( see e.g. [12], [19] and [20]). The results are shown for three different rings in Fig. 3(a), in Fig. 3(b) and in Fig. 3(c). The systems investigated are : Cr8 (s=5/2, $J$~17.2K ) [15]; Fe6(Na) (s=5/2, $J$~28.2K) [5]; Fe10 ( s=5/2, $J$~ 13.8K ) [5,19] (for last two compounds the respective complete formulas are : $[NaFe_6(\mu_2\text{-}OMe)_{12}(dbm)_6]Cl$ and $[Fe(OMe)_2(O_2CCH_2Cl)]_{10}$). In all three samples the ground state is non magnetic with total spin $S_{Total}=0$ and the energies of the lowest lying exchange multiplets can be described at first approximation by Lande's interval rule $E_S=2JS(S+1)/N$ where $N$ is the number of spins in the ring [5]. The main feature in the temperature dependence of $T_1^{-1}$ is the strong enhancement at low $T$ and the presence of a maximum at a temperature $T_0$ for each of the samples investigated. For $T<T_0$, $T_1^{-1}$ decreases approaching at low T an exponential drop due to the "condensation" into the $S_{total}=0$ singlet ground state as discussed later on. It should be noted that the behavior of the relaxation rate is different than the behavior of the uniform magnetic susceptibility. The latter, when plotted as $\chi T$, shows a continuous decrease with an exponential drop at very low temperature consistent with what expected for an AFM system with a singlet ground state. When the $T_1^{-1}$ is plotted together with $\chi T$ as shown in Fig. 3(a) for Cr8, one can see that the two quantities are approximately proportional in the whole temperature range except for the region where the peak in $T_1^{-1}$ occurs. In order to emphasize the critical enhancement it is more effective to plot the relaxation rate divided by $\chi T$ as shown in Fig. 4(a), Fig. 4(b) and Fig. 4(c) for the cases of Cr8, Fe6 and Fe10 respectively. As shown in the same figures the peak in the relaxation rate is depressed by the application of an external magnetic field and the position of the maximum moves at higher temperature on increasing the external field. One should also note the field dependence of the intensity of the maximum which decreases as the field is increased. In Fig.5 we report $(T_1^{-1})/(\chi T)$ vs. temperature for Fe6(Li), obtained by using 2 different nuclei, i.e. $^1H$ and $^7Li$. The qualitative temperature behavior below the peak is very similar for the two nuclei. Finally in Fig. 6(a) we plot the renormalized $(T_1^{-1})/(\chi T)$ data as a function of reduced temperature $t=T/T_0$ where $T_0$ is the temperature of the maximum for different samples and different fields. The sets of data overlap suggesting that the occurrence of a maximum is a universal effect of antiferromagnetically coupled rings with $S_{Total}=0$ ground state.

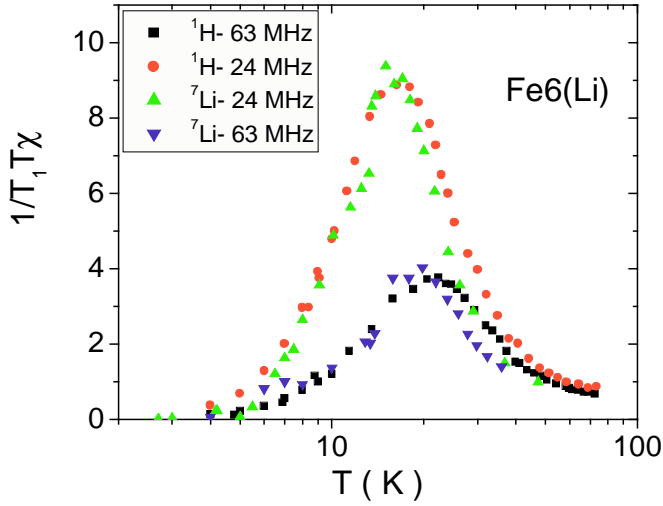

Fig. 5 Plot of renormalized $1/T_1 \chi T$ as a function of T for Fe6(Li), in two different magnetic fields for two different nuclei.

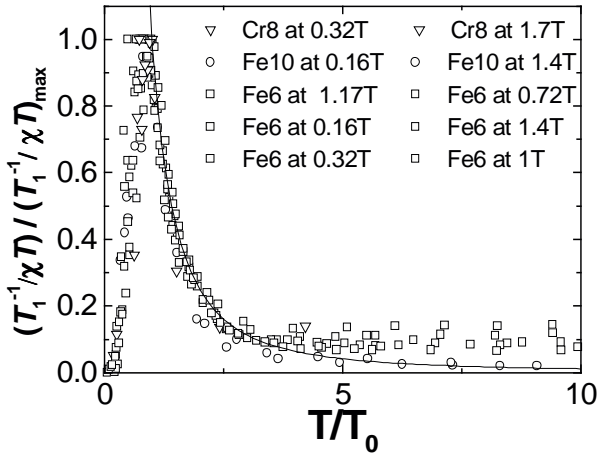

Fig. 6(a) : Plot of renormalized $1/T_1 \chi T$ as a function of $T/T_0$ for different magnetic fields and systems. All samples are in powders.

Recently it has been shown [21] that the relaxation rate data around the peak in Cr8 can be fitted very well by the simple expression for $T_1$ (Eq. 7) used in the high temperature approximation provided that the constant B is assumed to depend on temperature as $B = D T^n$ with the exponent n close to the value of three. The same kind of fit appears to work reasonably well also in Fe6(Na) and Fe10. This simple result is very surprising because it implies that the Eq. 10 which is obtained in the limit of high temperature remains valid even at low temperature provided that the spectral density of the fluctuations in Eq. 5 is allowed to narrow down and to become peaked at very low frequency. Thus from Eq. 10 and Eq. 5 one can write down an expression for the nuclear relaxation of the form:

$$1/(T_1 T \chi) = A' B/(1 + (H/B)^2) + C' \quad (msec^{-1}) \qquad (11)$$

which can be used to fit the data around the temperature of the peak. The constant C' in Eq. 11 group together the terms in Eq.10 which are weakly T dependent particularly in the region of the peak. The quality of the fit obtained using Eq.11 is exceptionally good for Cr8 as shown in the Fig. 4(a) and moderately good for Fe6(Na) and Fe10 as shown in Figs. 4(b) and 4(c). The constant B in Eq.11 can be identified either with $\Gamma_D/\gamma_e$ or with $\Gamma_A/\gamma_n$. In the first case the peak would be due to the slowing down of $\Gamma_D$ as $T^3$ from a value at room temperature much higher than $\omega_e$ to a value of the order of $\omega_e$ in the region of the peak. In the second case the peak would be due to the slowing down of $\Gamma_A$ from a value of the order of $\omega_e$ at room temperature (see previous paragraph) to a value of the order of $\omega_n$ in the region of the peak. Measurements of proton $T_1$ alone cannot distinguish between these situations. On the other hand measurements of relaxation on two nuclear species in the same molecular cluster can give us the answer. These measurements have been performed in the AFM ring Fe6(Li) in all identical to Fe6(Na) except for the replacement of Na with Li which induces a change of from J= 28K in Fe6(Na) to J = 23 K in Fe6(Li). The results are shown in Fig. 5. It is quite clear that the maxima in $T_1^{-1}$ overlap when the resonance frequency is the same and not the magnetic field. This is a direct proof that the constant B in Eq.11 has to be identified with $\Gamma_A/\gamma_n$ namely that the width of the Lorenzian in Eq. 5 becomes of order of $\omega_N$ at the temperature of the peak. Thus it appears that the peak arises from the spectral density of the longitudinal fluctuations $\Phi^z(\omega_n)$ in Eq. 10.

As a direct consequence of the fitting formula Eq. 11 (where C' can be neglected in the region of the peak ) one finds that the renormalized plot of $1/(T_1T\chi)$ vs $t=T/T_0$ shown in Fig. 6(a) has the simple form

$$1/(T_1T\chi) /(1/(T_1T\chi) )_{max} = 2 t^n /(1+t^{2n}) \qquad (12)$$

In Fig. 6(b), we show the renormalized plot compared to the function in Eq. 12. The fit is remarkably good with n=3 and no adjustable parameters.

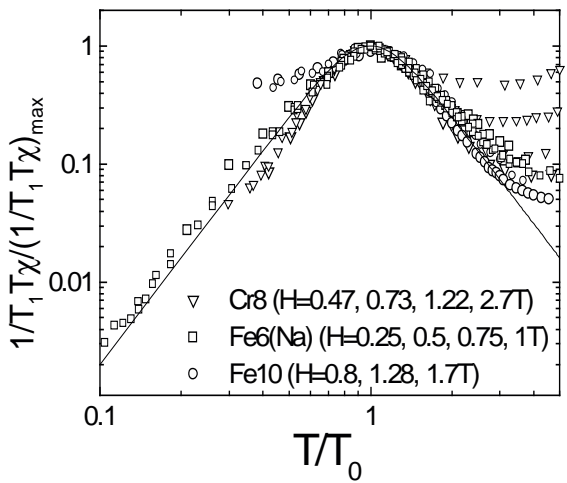

Fig. 6(b): Log-log plot of renormalized $1/(T_1\chi T)$ as a function of $T/T_0$. The solid line is theoretical curve by Eq. 12 with n=3. The samples are in form of powders.

At present there is no theoretical description which explains the critical enhancement manifested in the peak of $T_1^{-1}$ in AFM rings. One possible scenario is that the effect arises from the increase of the antiferromagnetic correlation which generates an enhancement and a slowing down of the spin fluctuations at the staggered wave vector $q=\pi$. To see this we rewrite Eq. 9 in an approximate form where we neglect any difference between the transverse (±) and longitudinal (z) terms and we divide the response function in a non critical part which is described by the $q=0$ term and a critical part described by the $q=\pi$ term:

$$1/T_1 = (h\gamma_n\gamma_e)^2/(4\pi g^2\mu_B^2)\, k_BT\, \chi(0)\, \{A(0)\, f_0(\omega_e)\, [\,1 + A(\pi)/A(0)\, f_\pi(\omega_e)/f_0(\omega_e)\, \chi(\pi)/\chi(0)] + A(0)\, f_0(\omega_n)\, [\,1 + A(\pi)/A(0)\, f_\pi(\omega_n)/f_0(\omega_n)\, \chi(\pi)/\chi(0)]\} \qquad (13)$$

By comparing Eq. 13 with Eq. 10 which fits so well the experimental data one would deduce that the critical term $f_\pi(\omega_e)/f_0(\omega_e)$ times, $\chi(\pi)/\chi(0)$ has a simple Lorenzian form with a correlation frequency $\Gamma_D = \gamma_e B$ which displays a critical behavior as $T^n$. This result is difficult to justify on the basis of dynamical scaling arguments similar to the ones used in phase transitions. First principles theoretical calculations of the spin correlation function could give the answer.

An alternative scenario is one in which the magnetic critical slowing down plays no relevant role in these finite size systems. In this case the peak in the nuclear relaxation rate could be simply related to a decrease of the cut-off frequency $\Gamma_A$ which reflects the anisotropy terms in the magnetic Hamiltonian which do not commute with the Heisenberg Hanıltonian. These terms determines the electronic spin lattice relaxation via spin-phonon coupling.

An important clue for the understanding of the problem is the dependence of the position in temperature of the peak of $T_1$ from the exchange interaction J in the AFM ring. The fits of the experimental data with Eq. 11 yields $B = D\, T^3$ with a value of D different for the three AFM rings. The values of the constant D are plotted in Fig. 7(a) vs the exchange constant J of the three AFM rings in a log-log plot. In Fig.7(b) the same values are plotted vs the energy gap $\Delta$ separating the ground state $S_T=0$ from the $S_T=1$ excited state according to Lande' rule $\Delta=4J/N$. In both cases there appears to be a negative power dependence i.e. $D \propto J^{-\alpha}$ and $D \propto \Delta^{-\beta}$ with $\alpha=4\pm0.5$ and $\beta=2.3\pm0.5$.

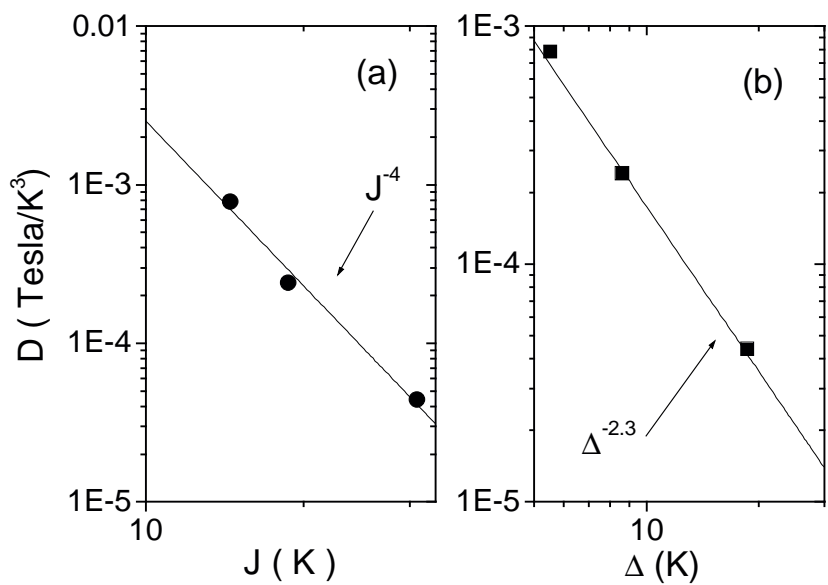

Fig. 7 : (a) A relation between D and J. (b) A relation between D and Δ.
The data refer to Cr8, Fe6(Na) and Fe10.

**b) Thermal fluctuations of the magnetization in nanomagnets: Mn12, Fe8**

The cluster containing twelve Mn ions, Mn12, has a high total spin (S=10) ground state in spite of the large antiferromagnetic coupling between the local Mn moments [14]. This circumstance together with the large magneto crystalline anisotropy generates an easy axis nanoferrimagnet with spectacular superparamagnetic effects [22]. Fe8 is similar to Mn12 with regards to the high spin ground state (S=10) but it has a more complex anisotropy and a lower barrier for the reorientation of the magnetization (about 27 K against the 67 K for Mn12) [23].The interest in molecular magnets is largely related to the possibility of observing quantum resonant tunneling of the magnetization (QTM) [24]. For the proper description of the quantum dynamical effects in the high spin ground state one has to take into account the environmental effects represented by spin-phonon coupling, intermolecular magnetic interactions and hyperfine interactions with the nuclei. Nuclear Magnetic Resonance is a very suitable local microscopic probe to investigate the above-mentioned environmental effects. The NMR spectrum at low temperature shows a multiplicity of spectral lines, which yield directly the local hyperfine field at the nuclear site due to the coupling with the local moment of the magnetic ions in the high total spin ground state. Furthermore the nuclear spin lattice relaxation is driven by the fluctuations of the orientation of the total magnetization of the molecular cluster and thus yields information about spin-phonon coupling which limits the lifetime of the m components (i.e. along the easy-axis) of the total spin S .

At very low temperature (≤ 4K) both Mn12 and Fe8 are in their magnetic ground state and the magnetization of the molecule is frozen in the time scale of an NMR experiment. Thus the nuclei experience a large static internal field. This allows to detect

$^{55}$Mn NMR in zero external field in Mn12 and $^{57}$Fe NMR in zero external field in Fe8. For protons the internal field is small being generated by the proton-Mn (Fe) moment dipolar interaction and thus the NMR in zero field is weak and can be observed only at low frequency over a broad frequency interval (2-4MHz) due to the presence of many inequivalent proton sites in the molecule. On the other hand the proton NMR in an external magnetic field shows a broad structured spectrum with a field independent shift of the lines of order of the internal field. For the analysis of the hyperfine field at the proton site and at the deuteron site in deuterated Fe8 as well as the temperature dependence of the spectrum we refer to ref. 25. At intermediate temperatures (4-30K) the magnetization of the Mn12 and Fe8 molecules are subject to large and fast fluctuations due to the thermal populations of the higher energy magnetic quantum states m= ± 9, ± 8, ± 7 … separated by crystal field anisotropy within the total spin ground state S=10. As a result the zero field NMR lines progressively disappear on increasing the temperature and the spin- lattice relaxation becomes very fast.

We have utilized this circumstance to investigate the thermal fluctuations of the magnetization by measuring the temperature and field dependence of $T_1^{-1}$. The study was possible for proton NMR [25] and μSR [26] in Fe8 over the whole temperature range. In Mn12 the proton relaxation becomes too fast in the temperature range (4-20K) and we had to measure the muon longitudinal relaxation by μSR technique [27]. For $^{55}$Mn NMR in Mn12 the signal can be detected only up to about 4K and for $^{57}$Fe NMR the signal is lost at about 1.8K because of a very short $T_2$. The proton $T_1$ vs temperature for Fe8 is shown in Fig. 8. The proton $T_1$ in Mn12 is shown in Fig. 9 vs magnetic field at low temperature below 4.2K. Regarding the temperature dependence in the range 4-30K we refer to the μSR relaxation rate since the proton NMR cannot be detected due to the short $T_1$. The results for the longitudinal muon relaxation rate ( a parameter analogous to $1/T_1$) are shown in Fig.10 for different applied longitudinal fields in Mn12 powders. The $^{55}$Mn $T_1$ vs temperature is shown in Fig. 11(b) in the narrow temperature range in which the signal is observable in zero external field. Finally the field dependence of the $^{55}$Mn $T_1$ is shown in Fig. 11(a).

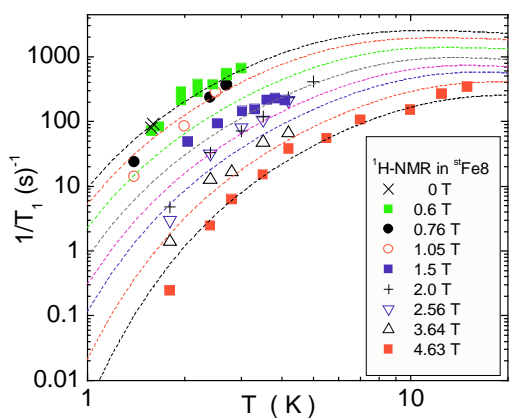

Fig. 8: Temperature dependence of $^1$H-$1/T_1$ in Fe8 "oriented" powders at different magnetic fields, parallel to the easy-axis. The lines are theoretical estimation calculated by equation 18 with a set of parameters described in the text.

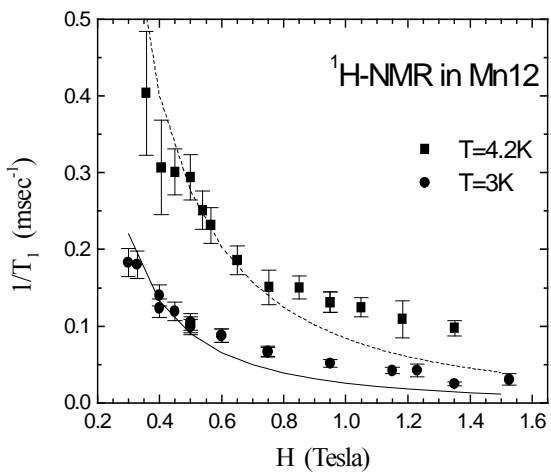

Fig. 9 : The external field (parallel to the easy-axis) dependence of $^1$H-$1/T_1$ in Mn12 "oriented" powders measured at T=4.2 and 3K. The curves are fitting results obtained by equation 18.

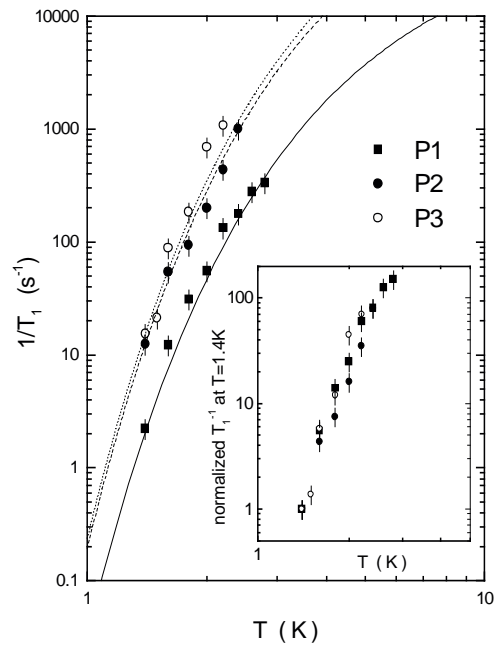

Fig. 11(a) : Temperature dependence of $^{55}$Mn-$1/T_1$ for each Mn sites in Mn12 oriented powders. Solid curves are fitting curves according to equation 18. The inset shows the temperature dependence of $1/T_1$ for the three peaks renormalized at the same value at T=1.4K.

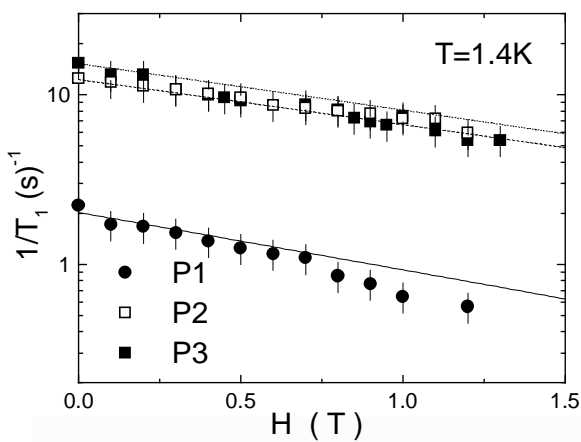

Fig. 11(b) : External field (parallel to the easy-axis) dependence of $^{55}$Mn-$1/T_1$ for each Mn sites in Mn12 oriented powders measured at T=1.4K. The solid lines are calculated results by equation 18.

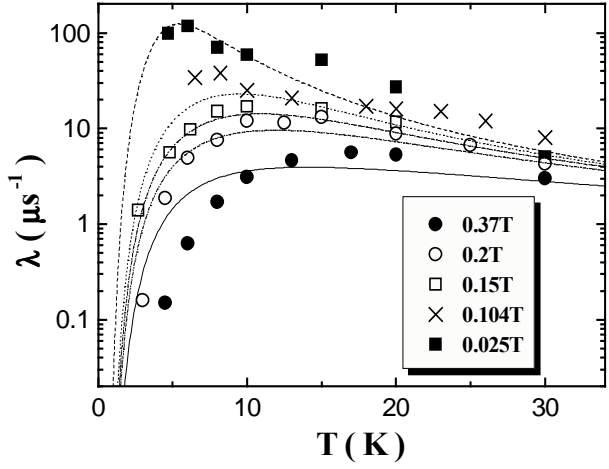

Fig.10. Muon longitudinal relaxation rate for different applied longitudinal fields in Mn12 powders, as a function of temperature. The lines are fits to the data following Eq. 18.

All the relaxation rate data shown here can be interpreted in terms of a simple model of nuclear spin lattice relaxation via a direct process driven by the fluctuations of the molecular magnetization. Since the fluctuations of the magnetization are a consequence of the finite lifetime of the $m$ magnetic substates due to spin-phonon interactions one can obtain from the fit of the NMR data the spin-phonon coupling constant, a quantity non easily derived by other techniques. We give in the following a brief account of the model. We start from a semi-classical approach by expressing the nuclear spin-lattice relaxation rate (NSLR) in terms of the correlation function of the transverse component $h_\pm(t)$ of the time dependent transverse hyperfine field at the proton site:

$$1/T_1 = \frac{1}{2} \gamma_N^2 \int <h_\pm(t)h_\pm(0)> \exp(i\omega_L t)dt \qquad (14)$$

with $\gamma_N$ the nuclear gyro magnetic ratio and $\omega_L$ the Larmor frequency. The fundamental assumption in this model is that, although the hyperfine field $h_\pm(t)$ is due to the interaction of the protons with the local moments of the $Fe^{3+}$ (Mn) ions, its time dependence is the same as the time dependence of the orientation of the total magnetization of the molecular cluster. Thus one can use for the correlation function of $h_\pm(t)$ an exponential function with correlation time $\tau_m$ determined by the lifetime broadening of the m sublevels:

$$<h_\pm(t)h_\pm(0)> = \sum_{m=+10}^{-10} <\Delta h_\pm^2> \exp\left(-\frac{t}{\tau_m}\right)\frac{\exp\left(-\frac{E_m}{KT}\right)}{Z} \qquad (15)$$

where Z is the partition function. The term $<\Delta h_\pm^2>$ is the average square of the change of the hyperfine field when the magnetization of the molecule changes orientation ( i.e. $\Delta m=\pm 1, \pm 2$). For sake of simplicity we assume an average value independent of the $\Delta m$ transition considered. The assumption should be acceptable since in the temperature range investigated most of the fluctuations occur between few m states close to the m=10 ground state. The lifetime $\tau_m$ for each individual m state is determined by the probability of a transition from m to m $\pm$ 1 , $W_{m,m\pm 1}$ , plus the probability for a transition with $\Delta m=\pm 2$, $W_{m,m\pm 2}$ i.e:

$$1/\tau_m = W_{m\to m+1} + W_{m\to m-1} + W_{m\to m+2} + W_{m\to m-2} \qquad (16)$$

The transition probabilities are due to spin-phonon interaction and can be expressed in terms of the energy level differences as [28,29]:

$$W_{m\to m\pm 1} = W_{\pm 1} = C' \, s_{\pm 1} \frac{(E_{m\pm 1} - E_m)^3}{\exp[\beta(E_{m\pm 1} - E_m)] - 1}$$

$$W_{m\to m\pm 2} = W_{\pm 2} = 1.06 \, C' \, s_{\pm 2} \frac{(E_{m\pm 2} - E_m)^3}{\exp[\beta(E_{m\pm 2} - E_m)] - 1} \qquad (17)$$

where

$$s_{\pm 1} = (s \mp m)(s \pm m + 1)(2m \pm 1)^2 \quad s_{\pm 2} = (s \mp m)(s \pm m + 1)(s \mp m - 1)(s \pm m + 2)$$

The spin-phonon parameter C' in Eq. 17 is given by C'= $D'^2$ /( $12\pi\rho v^5 h^4$) with $\rho$ the mass density  and v the sound velocity and D' a constant related to the crystal field anisotropy. Finally we can write for the NSLR:

$$1/T_1 = \frac{A}{Z} \sum_{m=+10}^{-10} \frac{\tau_m \exp\left(-\frac{E_m}{KT}\right)}{1 + \omega_L^2 \tau_m^2} \qquad (18)$$

where $A = \gamma_N^2 <\Delta h_\pm^2>$.

The energy levels $E_m$ in the above equations can be obtained from the Hamiltonian of the molecules expressed in terms of the total spin S:

$$H = -D\,S_z^2 - B\,S_z^4 + E\,(S_x^2 - S_y^2) + g\,\mu_B\,S_z\,H \qquad (19)$$

For Mn12 one has D=0.55 K, B=1.2x$10^{-3}$ K , and E=0 while for Fe8 one has D=0.27 K, B=0 and E= 0.046 K.

The experimental data both as a function of temperature at different fields (Figs. 8, 10 and 11(a)) and as a function of field at a fixed temperature (Figs. 9 and 11(b)) were fitted to Eq. 18 by using Eqs. 17 and 19 and treating  A and C' as adjustable parameters. From the fit of the proton relaxation in Fe8 one obtains the parameters : C' = 31 Hz/$K^3$ and A=

$1.02 \times 10^{12}$ (rad/sec)$^2$. The fit of the proton relaxation data in Mn12 (as well as the µSR relaxation data, see Fig.10) was obtained with a somewhat larger value of C' and a smaller value of $A = 0.45 \times 10^{12}$ (rad/sec)$^2$. The coupling constant A represents the average hyperfine interaction squared between protons and transition metal magnetic moments. The value found for Fe8 is larger than the value obtained in Mn12 indicating that in Fe8 the protons are subject to non negligible hyperfine interaction due to contact terms in addition to the dipolar interaction. From the knowledge of the spin-phonon coupling parameter C' one can estimate the lifetime of the m sublevels by using Eqs. 16 and 17. For a detailed discussion of the fitting parameters we refer to the original papers. Finally the $^{55}$Mn relaxation data were fitted with reasonable values of C' and A although for this case the two parameters cannot be determined independently. It should be mentioned that $^{55}$Mn relaxation data have also been reported by Goto et al. [30] and analyzed with a model based on a two state pulse fluctuation corresponding to the hyperfine fields in the two lowest energy levels m $= \pm 10, \pm 9$. This model is indeed a better model for the very low temperature region (< 4.2K) while our model has the advantage to be applicable even at higher temperatures where the higher energy levels become populated. A different approach has been proposed by Yamamoto and collaborator [31] based on a spin-wave approximation for the description of the energy levels and a Raman two magnon scattering mechanism to describe the $^{55}$Mn $T_1^{-1}$ data. Since both the phenomenological models based on a direct relaxation process [47,30] and the spin-wave model [31] appear to fit the low temperature $^{55}$Mn relaxation data a connection between the two approaches should be established.

# V. NMR at low temperatures ( $k_B T \ll J$)

When the temperature is much lower than the exchange interaction among magnetic moments in the molecule the system is mostly in its collective quantum ground state characterized by a total spin S. We have to distinguish the two cases of singlet ground state S=0 and of high spin ground state S>0. In the first case which pertains to AFM rings the residual weak magnetism of the molecule at low temperature is due to the thermal population of the first excited state which is normally a triplet S= 1 state. In the second case the molecule at low temperature behaves like a nanomagnet with a spontaneous magnetization proportional to the value of the ground state spin S. If there is no anisotropy the molecule acts like a soft nanomagnet with a magnetization which can be aligned by an external magnetic field with no hysteresis in the magnetization cycle. This is the case of Cu6 FM ring which is discussed in Section VI. In presence of an anisotropy the molecule behaves as a hard nanomagnet with hysteresis in the magnetization cycle. However, since there is no long range order each molecule acts as a superparamagnetic particle. At temperatures much lower than the anisotropy barrier the relaxation of the magnetization can be dominated by quantum tunneling. NMR can give interesting information in this low temperature regime. We will review the main results treating separately the case of a non magnetic ground state (AFM rings) and the case of a magnetic ground state (Mn12 and Fe8).

**a) Energy gap of AFM rings in the magnetic ground state**

AFM rings such as Fe10, Fe6, Cr8 already discussed in Section III, are characterized by a single nearest neighbor exchange interaction J which generates a singlet ground state of total spin S=0 separated by an energy gap $\Delta$ from the first excited triplet state S=1. From simple Lande's interval rule one has E(S)= 2J/N S(S+1). Thus in absence of crystal field anisotropy the gap is $\Delta$= 4J/N where N is the number of magnetic moments in the ring. In presence of crystal field anisotropy with axial symmetry characterized by the parameter D (see Eq. 19) the gap between S=0 and S=1, M=±1 is 4J/N +D/3 for the case of positive axial anisotropy. In Table II we summarize the magnetic parameters for the above mentioned three rings, for the Cu8 ring and for the cluster V12 which can be assimilated to a square of $V^{4+}$ magnetic ions.

From the inspection of the gap value in Table II it appears that below liquid helium temperature the molecular magnets will be mostly in the non magnetic ground state. In Cu8 the gap is so big that the ring is in its singlet ground state even up to room temperature. As a result the magnetic susceptibility goes to zero and the nuclear relaxation rate becomes also very small. Measurements of $T_1^{-1}$ in this low temperature range can yield interesting information about the energy gap $\Delta$ and about the quantum fluctuations in the S=0 ground state.

The electronic spin correlation function entering the expression of $T_1^{-1}$ ( see Eqs. 2 and 3) is defined as:

$$G_{ij}^\alpha (r,t) = \Sigma_n \Sigma_l \langle n|S_i^\alpha|l\rangle \exp(-\beta E_n + iE_n t/\hbar - iE_l t/\hbar) \langle l|S_j^\alpha|n\rangle \qquad (20)$$

where n,l number the eigenstates, $E_n$, $E_l$ are the energy eigenvalues, $\beta=1/k_BT$, $S_{i(j)}^\alpha$ are the spin operators of the $i^{th}$ ($j^{th}$) spin and $\alpha=x,y,z$. For a finite system the energy difference between eigenstates is very large compared to the Larmor Zeeman energy. Therefore for a direct process (see Eq. 3) only the matrix elements with n=l in Eq. 20 need to be considered and a broadening of the energy levels has to be introduced in order to fulfill energy conservation (i.e., in order to have some spectral density of the fluctuations at $\omega_e$ and $\omega_L$ in Eq.2). It should be noted that an alternative approach is to describe the nuclear relaxation in terms of a Raman process [31]. Even in this case one needs to have a broadening of the levels or a spin wave band. We have not explored this possibility since the direct relaxation process appears to be able to explain the experimental data.

TABLE II. Comparison of single ion spin values, exchange coupling constant (J) and energy gap $\Delta$(K) between the ground state and first excited state at H=0 for different AFM rings.

| AFM ring | magnetic ion spin | J (K) | $\Delta$(K)- no anisotropy |
|---|---|---|---|
| Fe10 | 5/2 | 13.8 | 5.5 |
| Fe6(Na) Trigonal | 5/2 | 28 | 18.7 |
| Cr8 | 3/2 | 17 | 8.5 |
| Cu8 | 1/2 | $\approx 1000$ | $\approx 500$ |
| V12 | 1/2 | 17.2 | 17.2 |

As a consequence of the presence of the Boltzman factors in Eq. 20 the NSLR at very low temperature will be simply proportional to the population of the excited states. For temperatures less than the energy gap $\Delta$ one has approximately:

$$T_1^{-1} = A \exp(-\Delta/k_BT)/(1+3\exp(-\Delta/k_BT)) \qquad (21)$$

where A is a fitting constant which contains the hyperfine coupling constants and the matrix elements in Eq. 20. The gap $\Delta$ depends on the applied magnetic field as a result of the Zeeman splitting of the excited triplet state of the AFM rings. The validity of the simple prediction of Eq. 21 was tested in the AFM rings described in Table II. The results of $T_1^{-1}$ vs T in Cr8, Fe6 and V12 are shown in Figs. 12(a), 12(b) and 12(c), respectively. For these systems the condition $T<<\Delta/k_B$ can be met in the temperature range (1.4-4.2K). For Fe10 one has to perform the measurements at lower temperature since the gap is smaller [32].

As seen in the figures the temperature dependence of $T_1^{-1}$ appears to be fitted well by Eq. 21 in the low T limit for Cr8 and Fe6 with values of the gap $\Delta$ in qualitative agreement with the values of the gap in TableII. Also the decrease of the measured gap with increasing field in Cr8 is in agreement with the notion that the application of an external field should close the gap. It should be noted that in presence of crystal field anisotropy the value of the gap is different from the one in Table II and, in presence of a magnetic field, the gap depends on the angle between H and the symmetry axis of the molecule. Thus measurements in powder sample cannot be expected to give better quantitative agreement than found here. The case of V12 is particularly interesting. In fact the low temperature proton relaxation data (see Fig. 12(c)) can be fit by the sum of a term described by Eq.18 with a gap value in agreement with Table II plus a constant term. In other words it appears that the relaxation rate saturates at low T instead of decreasing as expected for the "condensation" of the AFM ring in the singlet ground state. There are several reasons to exclude the possibility that the constant $T_1^{-1}$ at low T in V12 is due to paramagnetic impurities [33]. Then one can speculate that the low T contribution to $T_1^{-1}$ is due to quantum fluctuations in the singlet ground state for the quantum s=1/2 system (V12), fluctuations not evident for the classical s=3/2 (Cr8) and s=5/2 (Fe6) rings.

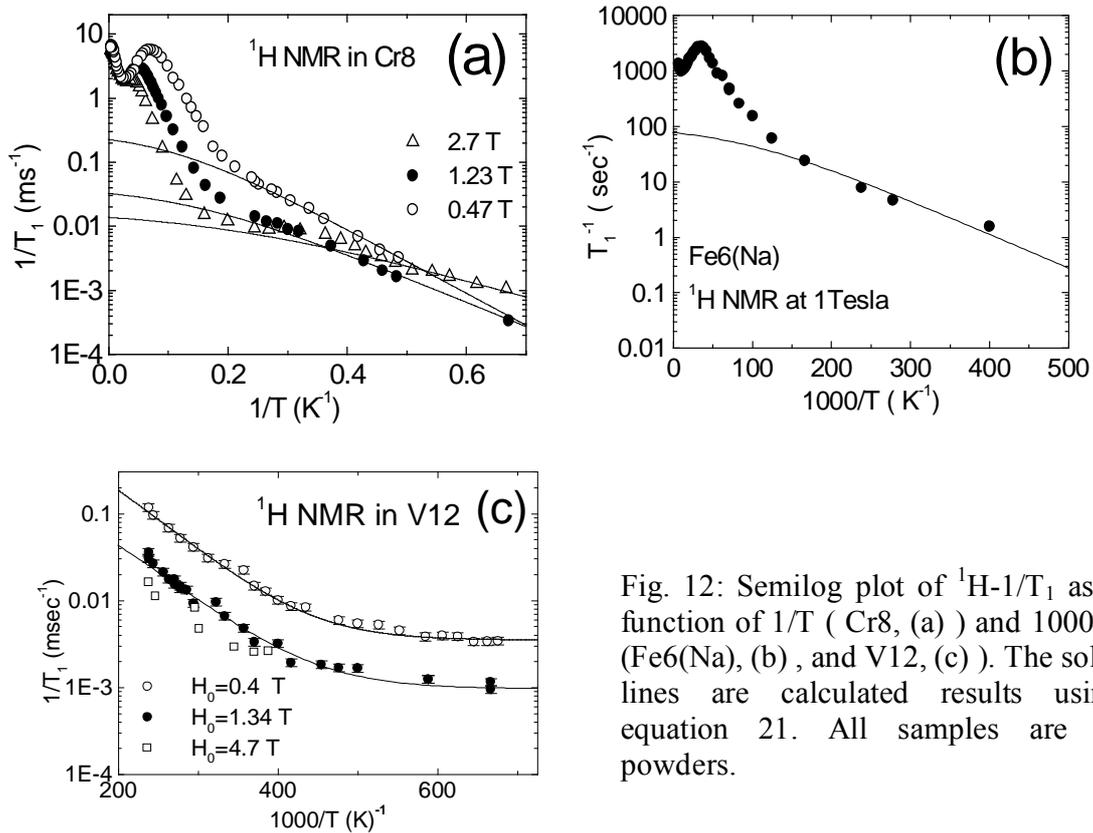

Fig. 12: Semilog plot of $^1$H-$1/T_1$ as a function of 1/T (Cr8, (a)) and 1000/T (Fe6(Na), (b), and V12, (c)). The solid lines are calculated results using equation 21. All samples are in powders.

The case of Cu8 appears to be the ideal case for the study of the energy gap and the singlet ground state since the gap is larger than room temperature. The Cu8 ring has been investigated both by proton NMR and $^{63,65}$Cu NMR and NQR [34]. The $^{63,65}$Cu spectrum is complicated by the presence of four non equivalent Cu sites. The $^{63,65}$Cu NQR spectra are composed of four separate lines for each isotope plus an additional line (probably due to impurities), spanning over the frequency range 16-22MHz. The $^{63,65}$Cu NMR spectrum in high field displays the powder pattern of a central line transition broadened by second order quadrupole effects (different for each of the four inequivalent sites) and anisotropic paramagnetic shift. On the other hand the proton NMR line is narrow ($\leq$ 40 KHz) and field independent. The nuclear spin-lattice relaxation rate (NSLR) decreases exponentially on decreasing temperature for all nuclei investigated as shown in Fig. 13 and Fig. 14. In both figures we have drawn the curves corresponding to Eq. 21 with the gap value given in Table II and derived from susceptibility measurements. One can see that both for $^1$H NMR and $^{63,65}$Cu NQR-NMR the experimental NSLR deviates from the exponential behavior at relatively high temperature in a manner similar to the case of V12 discussed above. However, in the case of Cu8 it appears more difficult to rule out the effect of paramagnetic defects and thus it is still to be proved that the NSLR in rings of spin 1/2 in the singlet ground state is driven by quantum fluctuations.

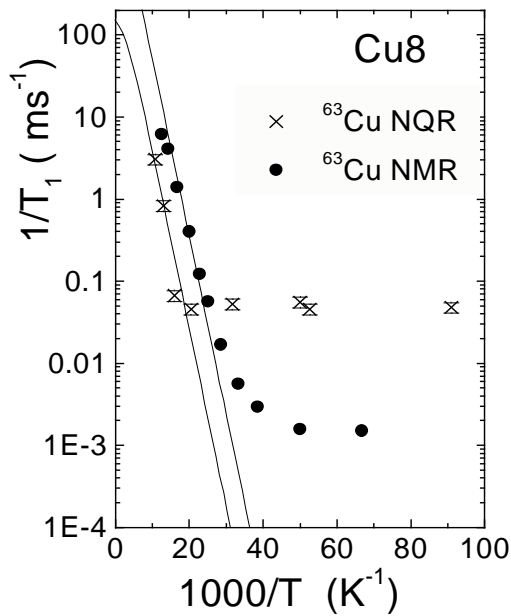

Fig. 13 : Semilog plot of of $^{63}$Cu-$1/T_1$ as a function of 1000/T in Cu8 powders.

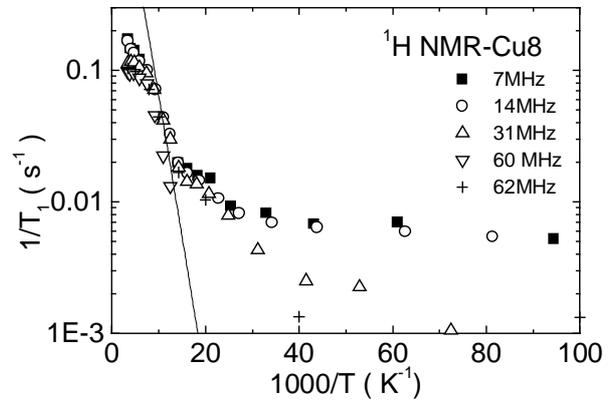

Fig. 14 : Semilog plot of of $^1$H-$1/T_1$ as a function of 1000/T in Cu8 powders.